\documentclass[10pt,twoside]{article}

\usepackage{asp2004}
\usepackage{epsf}
\usepackage{psfig}
\usepackage{lscape}
\usepackage{natbib}

\begin{document}
\title{Photometric and Astrometric Calibration of the JWST Instrument Complement}   %%% Fill in title
\author{Rosa I. Diaz-Miller}   %%% Fill in author names
\affil{Space Telescope Science Institute, 3700 San Martin Dr., Baltimore, MD 21218, USA}    %%% Fill in author affiliations
%\keywords{techniques: photometric, astrometry, infrared: stars }

\begin{abstract} %%% Abstract to run on from here.
In preparation for James Webb Space Telescope (JWST), a set of cross calibration programs with HST and Spitzer for suitable primary photometric standards and astrometric fields were developed.  NICMOS/HST and IRAC/Spitzer photometry observations of new solar analog standards in NGC\ 6791 and Melotte\ 66 were executed. These new photometric standards will provide $\sim 5$\% photometric precision at V~$\sim$~19 from the near-IR to the mid-IR wavelength range for efficient on-orbit calibration and measuring of photometric stability of the JWST complement.  For the astrometric calibration, a field in the LMC has been selected. This field falls within the JWST continuous viewing zone, within 5\deg from the ecliptic poles, and has the stellar density necessary to achieve accuracies better than 1 mas with HST/ACS. These independent observations will play a key role in meeting the mission requirements and will allow a fast commissioning of the observatory.
\end{abstract}

%%% MAIN BODY OF TEXT GOES HERE. CONSULT "INSTRUCTIONS FOR AUTHORS USING
%%% LATEX2E MARKUP", SECTIONS 2.3-2.6 FOR HELP WITH EQUATIONS, FIGURES,
%%% AND TABLES.

\section{Introduction} 

The James Webb Space Telescope (JWST), led by NASA's Goddard Space Flight Center  in collaboration with the European Space Agency (ESA) and the Canadian Space Agency (CSA), is a 6-M class observatory optimized for the near and mid IR. It has
a  three-mirror anastigmatic design with a primary constructed from 18 hexagonal deployable berylium mirror segments. A large sunshield will protect the Optical Telescope Element (OTE) and instruments from the sun and earth radiation and will enable passive cooling of the telescope and instruments. The imaging and spectroscopic capabilities will span the 0.6-28 $\mu m$ region with a diffraction-limited image quality at 2 and 4 $\mu m$. To minimize station-keeping maneuvers, the observatory will have a second Lagrange point (L2) orbit with a period of about 6 months about the L2 point \citep{Stal04}.

The main scientific goals of the observatory span many orders in
time, space, and size, from the first light objects, reonization, and the assembly of galaxies  to the
birth of stars, proto-planetary systems, planetary systems and the origins of life \citep{Gaal06}.
JWST will address these themes with its four main scientific instruments:
the Near Infrared Camera (NIRcam), the Near Infrared Spectrograph (NIRSpec), the near-infrared Tunable Filter Imager (TFI), and the Mid Infrared Instrument (MIRI). Table \ref{tab-1} summarizes the main instrument characteristics.

\begin{table}
\caption{Science Instrument Characteristics \label{tab-1}}
\smallskip
{\footnotesize
\begin{tabular}{lccccc}
\tableline
\noalign{\smallskip}
Instrument & Wavelength & Optical Elements & FOV & Plate Scale & Pixels Format\\
 & ($\mu m$) & & & (mas/pixel)  \\
\noalign{\smallskip}
\tableline
\tableline
\noalign{\smallskip}
NIRcam & 0.6-2.3 &fixed filters& 2\farcm2 x4\farcm4 & 34	& Two 4096$^2$ \\	
  (Short Wave) &  &  R$\sim$ 4, 10,100 & &	& \\		
&  & coronagraphic spots  & & & \\
  (Long Wave)& 2.3-5    & & & 68 & Two 2048$^2$ \\
\noalign{\smallskip}
\tableline
\noalign{\smallskip}
NIRspec & 0.6-5 & Grating (R=1000) & 3\farcm4x3\farcm1 & 100& Two 2048$^2$ \\
 &  & Prisim (R=100) & &  & \\
 &  & slits, MSA & &  & \\
 \tableline
 \noalign{\smallskip}
 MIRI & 5-27 & filters &1\farcm4x1\farcm9 & 110 &1024$^2$   \\
(imaging) &  & coronagraphic spots &(26\arcsec x26\arcsec & &   \\
&  &  phase masks &  Chor.)& &   \\
 (spectroscopic) &   5-10 & prism (R $\sim$ 100) & & & \\
  &5-27 &  IFU (R$\sim$ 3000) & 3\farcs6x3\farcs6 to& 200 to &  Two 1024$^2$  \\
 &           &                                        & 7\farcs5x7\farcs5 & 470 &    \\
 \tableline
  \noalign{\smallskip}
FGS-TF &1.2-4.5 & filters + & 2\farcm3x2\farcm3 & 68 & 2048$^2$ \\
& &  etalon (R$\sim$100) & & & \\
\noalign{\smallskip}
\tableline
\end{tabular}
}
\end{table}

To accomplish these objectives, we have designed a set of cross calibration programs with
the HST and the Spitzer observatories. These will provide accurate absolute
photometric and spectrophotometric calibrators, as well as the required observatory astrometric accuracies.
This paper presents the motivation behind the current  cross-calibration programs and sumarizes the targets and
the strategy followed in order to achieve the required accuracies. Here, we discuss only the
photometric and astrometry cross-calibration programs. For the details of the spectrophotometric cross-calibration, please refer to the paper of R. Bohlin elsewhere in these proceedings.

\section{Rationale for cross-calibration of JWST}

To enable the scientific goals of JWST, it is necessary to achieve high accuracy on the calibrations.
The calibration requirements are outlined in Table \ref{tab-2}

\begin{table}
\caption{JWST Required Calibration Accuracy \label{tab-2}}
\smallskip
\begin{center}
{\small
\begin{tabular}{lcccc}
\tableline
\noalign{\smallskip}
Instrument & Flux (\%)& Flux (\%) & Flux (\%) &Wavelength (\%) \\
 &  Imaging &  Coronagraphic& Spectroscopy & (resolution elements)\\
 & &  Imaging & &  Spectroscopy\\
 \noalign{\smallskip}
 \tableline
 \noalign{\smallskip}
NIRcam &	5&	5&	N/A&	N/A\\
NIRSpec	&N/A&	N/A	&10	&12.5\\
MIRI	& 5	&15	&15&	10\\
FGS-TF&	5	&10&	N/A&	10\\
\noalign{\smallskip}
\tableline
\end{tabular}
}
\end{center}
\end{table}

 \noindent
 Furthermore, JWST should be capable of making relative offsets and pointing with an astrometric accuracy of 5 milliarcsec RMS.

In order to derive a set of accurate and consistent photometric and astrometric calibrators, we 
take advantage of the space based observatories HST and Spitzer. The high stability of these observatories is pivotal to building a baseline calibration for JWST and to providing a solid basis for the astrometric and photometric calibrations. These observations will have the added advantage of making the HST calibration more robust and enable  cross-calibration of HST, Spitzer, and JWST. In particular, HST astrometric observations  are key in meeting JWST calibration requirements early in the lifetime of the mission and to the accuracy level required  because the JWST  architecture allows rolls of not more than
$\sim5^{\circ}$ at any one time. As we will discuss later, multiple roll angles are
required for accurate astrometric calibration. 
 Due to the finite lifetime of HST and Spitzer we are currently conducting the JWST calibration programs in collaboration with a large number of people, both within STScI and elsewhere (see acknowledgements).

\section{Photometric Calibration}

\begin{figure}
\begin{center}
\plotfiddle{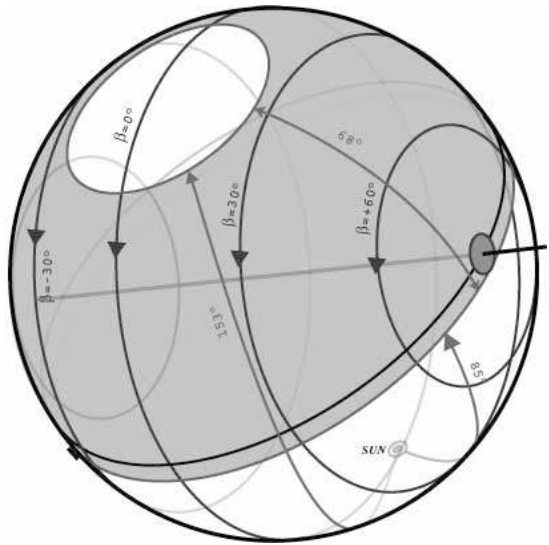}{8cm}{0.0}{50}{50}{-100}{0}
\caption{JWST instantaneous FOR is shown in grey. The FOR is limited by the size of the sunshield. The maximum circle of the sphere intercepting the sun is the ecliptic. The CVZ is shown with the darker gray. \label{DM1}}
\end{center}
\end{figure}

The photometric cross-calibration program will provide stellar sources to determine the absolute calibration for JWST photometric bandpases. The high sensitivities of the JWST instrumentation pose limitations on the brightness of the standards stars that can be used to derive the absolute photometry, and these are further constrained by the minimum integration time for NIRcam full field observations (10 seconds). The limit is set to K =15, below which the detector saturates. Most of the current IR standard stars are brighter and, therefore, cannot be used to calibrate JWST. Furthermore, a large number of these are  in the ecliptic plane where, due to JWST observing restrictions, targets become accessible only for approximately 69 continuous days twice a year.  Using a large system of ground-based and space-borne observatories, Cohen et al. (1999) constructed an all-sky network of stellar standards; but again these stars are still above the limits of JWST detectors; and therefore, it is necessary to look for more suitable standard stars. Besides the brightness constraints, JWST standards should also be readily accessible to JWST; i.e. in regions of the celestial sphere where the telescope can point at any time.  The accessible regions for JWST are determined by the size of the sun shield and the position of the sun. Figure \ref{DM1} shows  an instantaneous Field of Regard (FOR) for JWST. The gray area indicates the region in the celestial sphere that  JWST could observe at a  given time. In particular, the dark grey area indicates the  continuous viewing zone (CVZ), which is accessible to the observatory throughout the year.

The absolute photometric cross-calibration program for JWST,  is based on the solar-analog method (Johnson 1965; Campins, Rieke, \& Lebofsky 1985), which has been used also as an independent validation of the absolute calibration of infrared (IR) photometry for Spitzer's Multi-band Imaging Photometer (MIPS; Rieke et al. 2006). The solar-analog method relies on the fact that solar analog stars have a small number  of strong absorption features at IR wavelengths and, therefore, their spectral energy distribution (SED) can be well represented by the solar flux. In this case, the solar spectrum or a model is used to fill the gaps between ground based and space based observations after it has been normalized to the V magnitude of the solar analog. Although variations can be expected on the flux distribution due to the selection criteria of the solar-analog stars (Hardorp 1978; Wamsteker 1981),  this method leads to accuracies between 3\% and 5\% for photometry in the J, H, K, L and M bands (Campins, Rieke, \& Lebofsky 1985). Lower accuracies are found at longer wavelengths, where Rieke, Lebofsky, \& Low (1985) determine that the absolute calibration accuracy at 10 and 20 $\mu m$ is only about 3\% and 8\%, respectively.  

The NIRcam science team identified a set of clusters suitable for JWST photometric calibration purposes.  The appropriate clusters were chosen due to their stellar magnitude range (K = 15.3 and 16.9), their location on the FOR, and their known low extinction values. Three of these clusters, NGC~6791, NGC~2420, and NGC~2506, have been thoroughly studied and a large amount of information exists about them. The fourth cluster,  Melotte 66, is not as
well know; however, it has a favorable positioning in the FOR, closer to the CVZ, and therefore most likely to be readily available for observation at launch time. Another consideration in targeting clusters, rather than isolated stars, had to do with efficiency of the observation and eventually of the quality of the calibrations. Selecting stars within a cluster reduces the differences in our sample, i.e., we can assume that  all the members of the a given cluster have
uniform metallicity, same distance, age, and extinction. Table \ref{tab-3} gives a list of these parameters for the above mentioned clusters.  Another advantage of selecting clusters is that more than one star can be observed in any one orbit. Given the limited amount of observing time we can secure with HST, we need to make sure that  the available observations will give us the information and accuracy required. Having several stars in one cluster can also improve our ability to identify  systematics between the predicted and observed fluxes and, together with the knowledge of the properties of the clusters, will serve to better characterize the models.

\begin{table}
\caption{Properties of candidate JWST photometric calibration clusters \label{tab-3}}
\smallskip
\begin{center}
{\small
\begin{tabular}{lccc}
\tableline
\noalign{\smallskip}
Star Cluster &E(B-V)& Fe/H & m-M  \\
 \noalign{\smallskip}
 \tableline
 \noalign{\smallskip}
NGC~2420 & 0.04 & -0.4 & 12.0\\
NCG ~2506 & 0.05 & $\sim$ -0.3 & 12.5\\
NGC~6791 & 0.13 & +0.3 & 13.4 \\
Melotte~66 & 0.14 & -0.6  & 13.75 \\
\noalign{\smallskip}
\tableline
\end{tabular}
}
\end{center}
\end{table}
      
 The objective of this program  is to derive accurate zero-point fluxes for JWST instruments, in particular for NIRcam. The accuracy with which we can predict these fluxes depends strongly on the accuracy of the observed solar spectra used to model the observations (e.g. Thuillier et al. 2003; Kurucz 2005a) via the solar-analog method, as well as on the accuracy of the photometric data.  In the later case, the accuracy of these data will determine our ability to validate the suitability of the selected stars as standards. In order to constrain the models, it is imperative to secure data at different wavelength bands and therefore, close collaboration between the Spitzer Science Team, the HST Mission, and the JWST Mission is key to the success of this program. ln order to further validate our  estimates, we will make a detailed comparison of our  data with several stellar atmosphere models (e.g. Kurucz 2005b, 2006; Lejeune \& Cuisiner 1997). In this case, the spectral type of the candidate stars will be confirmed  by means of optical photometric observations (e.g. Stetson, Bruntt, \& Grundahl 2003). Using the knowledge of the physical properties of the cluster, as well as assumptions on the Spectral Type-Teff relations of
de Jager \& Nieuwenhuijzen (1987) the best atmosphere models for these solar candidates will be identified. This
in turn will be used to compare with the observed flux for the HST/NICMOS and Spitzer/IRAC bandpasses, as well as for the optical data. The observed/predicted residuals can then be used
to further constrain the spectral types and the suitability of the selected stars.  Comparison of the prediction to observed fluxes for all the members of any one cluster  will help us to further constrain the models and observed solar spectra as well as to search for any systematic trends.
 
The HST/JWST cross calibration program includes IR observations with the HST/ NICMOS NIC2 camera using  F110W, F160W, and F205W filters (centered at $\lambda $ = 1.12, 1.6, and 2.066 $\mu m$ respectively).
These will be complemented with  NIR IRAC Spitzer observations at 3.6, 4.5, 5.8, and 8 $\mu m$ and if possible with MIPS at 10 and 24 $\mu m$.
Currently, we have already secured NICMOS observations of six stars: three are members of NGC~6791, and the other three were observed in Melotte~66; the J2000 coordinates of the selected targets are listed in Table \ref{tab-4}. For NGC~6791 the calibration standard stars  were selected using $B-V$ color derived Stetson, Bruntt, \& Grundahl  (2003). Column 4 of Table \ref{tab-4} gives the $B-V$ color for these stars. For Melotte 66, on the other hand, we used the $V-I$ colors determined by Kassis et al. (1995).
For both clusters, the stars were selected according to their locus with respect to other stars within the cluster.  
Given the small plate scale of HST/NICMOS NIC2 camera (0\farcs075/pixel) a single observation is limited to only small regions within the cluster; therefore, careful selection of the observing field has to be made to accommodate the large plate scale of Spitzer/IRAC camera (1\farcs2/pixel) in order for it to produce an accurate measurement of the flux of the stellar source. We, therefore, limited our HST/NICMOS observations to stars isolated enough for Spitzer/IRAC to resolve them. It is also due to this constraint that we are still in the process of identifying the candidate stars for NGC~2420 and NGC~2506 for our HST/NICMOS observations, while
IRAC has already completed observations for NGC6791, NGC2420, and NGC2506. In the future we hope to be able to secure Spitzer observations of Melotte~66. 

\begin{table}
\caption{JWST cross-calibration photometric targets \label{tab-4}}
\smallskip
\begin{center}
{\small
\begin{tabular}{lllllllc}
\tableline
\noalign{\smallskip}
Star Cluster & \multicolumn{3}{c}{RA}& \multicolumn{3}{c}{DEC} & color\\
 \noalign{\smallskip}
 \tableline
 \noalign{\smallskip}
NGC6791 & 19h & 21m &4.45s & +37\deg & 47\arcmin & 54\farcs7 & 0.9 \\
 & 19h & 20m & 55.53s & +37\deg & 47\arcmin & 39\farcs67 & 0.902\\
 &19h & 20m & 51.33s &+37\deg & 48\arcmin & 41\farcs03  & 0.889\\
  \tableline
 \noalign{\smallskip}
Melotte66 \tablenotemark{a}& 07h & 25m & 57.07s & -47\deg &  34\arcmin & 31\farcs74  & 0.84\\
& 07h & 26m & 5.28s & -47\deg & 35\arcmin & 7\farcs16 & 0.88  \\
(2 stars) & 07h & 26m & 11.52s & -47\deg & 35\arcmin & 9\farcs6 & 0.82 \\
\noalign{\smallskip}
\tableline
\end{tabular}
}
\end{center}
\tablenotetext{a}{The coordinates for these stars were determined using the charts in Anthony-Twarog, Twarog, \& Sheeran (1994)} 
\end{table}

\section{Astrometric Calibration}

Due to the small apertures used by the JWST/NIRSpec multi-object spectrograph, a high accuracy on the astrometry should be achieved so that we can position the observed targets within the
desired window. Current science requirements of JWST call for a distortion correction within any instrument and guide star to not exceed 0.005 arcsec RMS. If we take into account that some error is inherent in the pointing and slew of the spacecraft, the accuracy of the astrometry should be less than or about 1 mas. JWST calibrations alone cannot achieve such an accuracy. The design of the observatory limits its roll angles to  $\le \pm 5\deg $ at any one time. As is going to be discussed below, in order to derive accurate geometric corrections for any astrometric field, observations should be carried out at two different angles. Unfortunately, the roll constraints make this strategy unfeasible for JWST because it imposes a wait period of more than 60 days before a good self astrometric calibration can be done.  This might be longer than the expected stability times for JWST's  multi-mirror figure.
We, therefore, have to resort to the only other way of doing distortion calibration: creating an
astrometric-standard field with another instrument, in this case HST/ACS, with which we can obtain the adequate observations and with which we can reach the desired accuracies. 

A great deal of work has been devoted to finding the geometric distortion solutions for HST/ACS instrumentation (Anderson \& King 2002, 2004, 2006). Using the PSF fitting method developed by Anderson \& King (2006; hereafter  AK06), which relies on a well modeled effective PSF library at different locations of the detector, current  ACS/WFC images can achieve  astrometric accuracies of 0.01 pixels. This would correspond to 5 mas; taking a plate scale for ACS/WFC to be 0.05 arcsec.  This library is composed of an array of 9 x 10 fiducial PSFs from which the PSF for a particular star in another position can be interpolated. The size of this PSF array was determined by the variations in the charge diffusion and optical aberrations with position on the ACS/ WFC detector. 

There are two factors that are key for accurate astrometry: the ability to determine the position of the stars  and the ability to  correct for geometric distortion. The accuracy reached for the position of the stars depends strongly on the careful treatment of the PSF, while the correction for distortion relies on knowledge of the types of distortion that affect the instrument and our ability to model them correctly.  All the optical instruments present some amount of geometric distortion that results in changes of the true point source position on a detector. In order to be able to correct for this distortion, we must have
a good set of exposures taken at different pointings so that we
can find the single distortion solution that will allow the positions
in the different frames to be related by simple offsets and rotations.
The distortion correction is usually a non-linear correction
to account for high-order perturbations caused by the optics; however
as Anderson \& King (2003) noted for the WFPC2, there is also a linear 
component that contributes to distortion.  It can be thought of as a lack 
of perpendicularity between the axes or as a difference between the scale
of the x and y axes.  This is quite evident for HST/ACS instruments, for which the x axis is inclined by  82\deg with respect to the y axis, resulting in a rhomboid-shaped field. In order to correct for this skew term of the distortion, we need  to secure data at different  offsets and also at different rotations. Furthermore, Anderson \& King (2002) found that there is some component of the ACS/WFC distortion that is introduced by the filters and that varies from filter to filter. It is clear that  in order to achieve the highest accuracy possible with ACS/WFC, we must carefully plan our observations. 

But what constitutes a good astrometric field?
First of all, a good astrometric field  should have a large number of bright, but not saturated, stars that can be used to derive a well sampled effective PSF and to model all scales of the distortion. It should be large enough to cover the FOV of all JWST instruments ( $>$ 4\farcm4; JWST-OPS-002843). The selected field should also have a uniform stellar density distribution, with variations from region to region of less than a factor of two. Since the objective of astrometry is to determine accurately  the star positions,  most of the stars within the field should be isolated enough so as not to compromise our ability to get accurate and repeatable positions. The typical  separations for stars that can be measured well is about 10 FWHM or so (J. Anderson, private communication). Another important factor to consider is the luminosity function. Ideally, we would like to have a field where all the stars have uniform brightness where we can get plenty of well measured stars, with S/N of about 100 or better, but not many saturated stars so as to make much of the field unusable. Small or negligible proper motions are also desirable, as we want a field that does not change with time.  
Further considerations have to be made in the choice of filters in order to construct an astrometric-standard field that is  also good for the JWST instruments working in the NIR to MIR wavelengths ranges.
    
 The high sensitivity of JWST instruments poses a limit on the brightness of the stars at IR wavelengths ranges. In the case of JWST, stars with magnitude K~$>$~15 can  be used to produce  useful fields of unsaturated stars. This magnitude would correspond to a magnitude  V~$>$~15, 17, 21, and 22  for A2V, G5II, M7V, and M6III stars, respectively (Zombeck 1990). Taking these limits into account, we 
pose the extra requirement that the ideal field provides a considerable number of stars with V=24 that can be observed with the ACS/WFC instrument, with a  S/N~$\sim$~100.  Furthermore, in order to determine time-dependent effects of the observatory that could affect the distortion, the astrometric-standard field should be within the CVZ, so it is available at any time.  
 
 \begin{figure}
%\plotfiddle{diaz-miller_fig2.eps}{6cm}{0}{80}{80}{-100}{0}
\plottwo{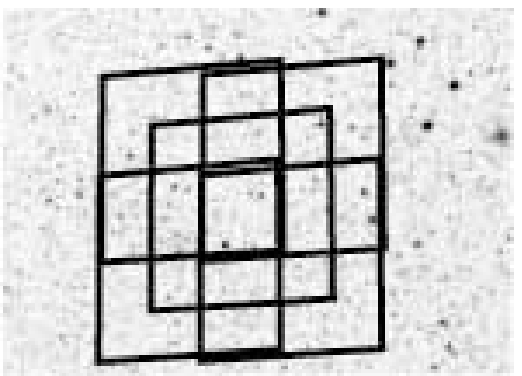}{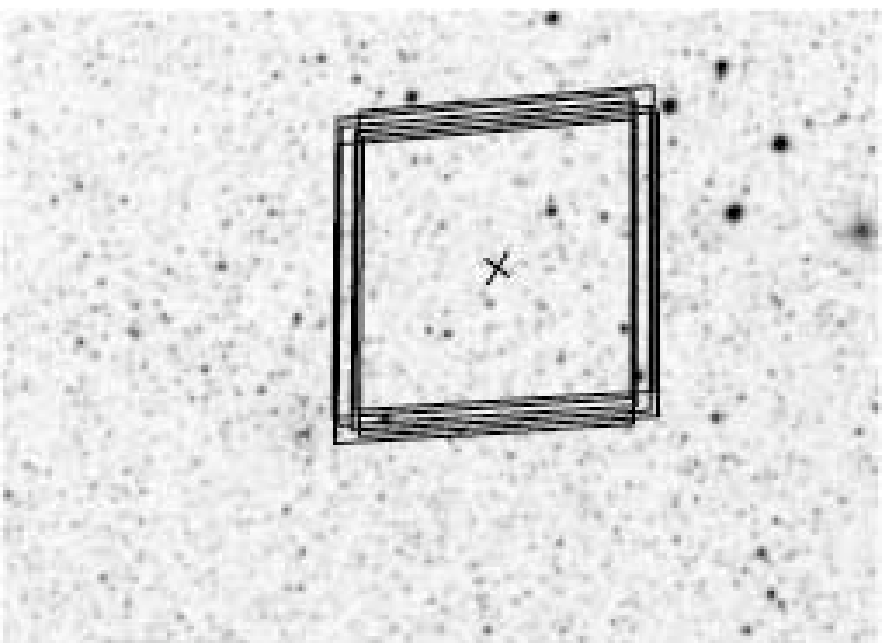}
\caption{JWST Astrometric-standard field in the LMC. Left: the different
pointings (one per orbit) obtained with ACS/WFC camera for the first
orientation. Each rectangle encompases the area covered by all the dithering
positions in one orbit. Right: Dithering positions for the top right pointing
shown in the left panel. Each of the 6 rectangles shows the FOV of ACS/WFC.\label{fig-2}}
\end{figure}

 The first target that comes to mind when looking for high stellar density within the CVZ is the  Large Magellanic Cloud (LMC). In order to determine  the stellar density distribution of this field we use the   Zaritsky et al. (2004) catalog which, although complete up to V=21 magnitude, gives us a good estimate of the stellar density that can be expected at lower magnitudes. For comparison, we also look for ACS/WFC observations in the LMC. After analyzing data from ACS/WFC programs 12001 (PI J. Green) and 9891 (PI G. Gilmore ),  and comparing the stellar density of these regions with those of the Zaritsky catalog, we identified a potential astrometric field centered at RA 05h 22m 40s and DEC -69\deg 32' (J2000).  This field is at the North edge of the LMC bar  and shows no bright stars that could saturate either ACS/WFC or JWST instruments. It is also far from any of the region where gas emission or dust has been observed.
In order to adequately sample this field for astrometric purposes, we secured 7 orbits with ACS/WFC. Five of these orbits are used to form an  X pattern in the sky  that covers a  5\farcm4 x 5\farcm4 field of view. Each of these orbits is taken at the same orientation but with a different pointing that allows an overlap in the observed region by about three quarters of the ACS/WFC detector. A complete sample  is secured via further dithering by steps of about 2\arcsec to 10\arcsec for each of these pointings, for a total of 35 different exposures in the first orientation and 14 in the second.  Figure \ref{fig-2} 
shows the FOV for each of the offsets super-imposed on the STScI DSS image of the region. Also, this figure shows the dither pattern for one of these pointings. In this case, the last two orbits were planned so that they were rotated from the original orientation by about 75\deg to 105\deg. For these,  the dither pattern is loosened by selecting dither steps of about 45\arcsec while keeping the pointings only about 1 arsecond apart. The resulting mapping of the region covers almost the entire observed field in the primary orientation. For the sake of completeness, we also added one short exposure at the beginning of each visit. This will give us a larger dynamical range over the region and the possibility to correct for the position of saturated stars.  

\begin{figure}
%\plotfiddle{diaz-miller_fig4.eps}{20cm}{0}{80}{80}{-100}{0}
\plotone{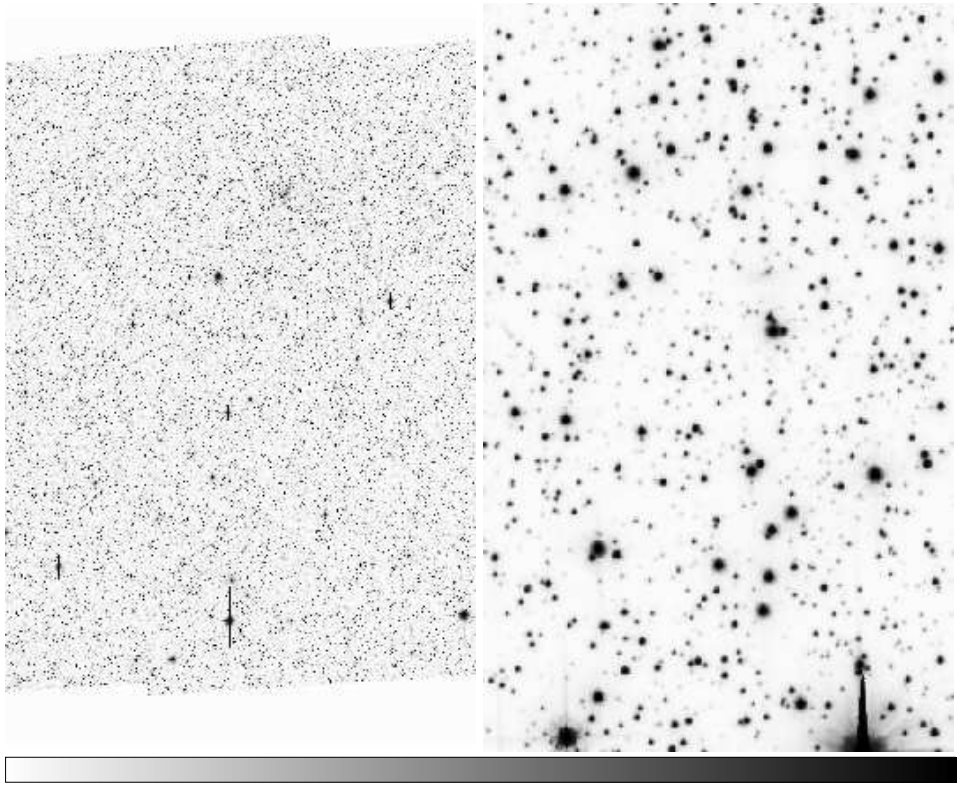}
\caption{{JWST astrometric-standard field in the LMC. The left image, centered at  RA 05h 22m 40s and DEC -69\deg 32\arcmin (J2000), shows our field composed of 30 343 sec ACS/WFC exposures obtained at the same orientation. A close up of the same field centered at the same position, is shown in the right side image.}\label{fig-4}}
\end{figure}

Although the data have not been fully analyzed, a preliminary evaluation shows that there are of the order of 100,000 stars in the field that are good for astrometric analysis. Figure \ref{fig-4} shows a stack of the 30 images obtained in the first five orbits at the same orientations (J. Anderson). This image shows an excellent field with uniform stellar brightness distribution. Astrometric accuracies should be better than 0.01 ACS/WFC pixels. This, however, will involve more work than we expected.
As we mentioned before, AK06 derived a PSF library for most of the ACS/WFC filters that will fit stellar images with a centroid accuracy of 0.01 pixel.  However, these images show a considerable focus change that might warrant  a new effective PSF library constructed for this particular region. Fortunately, J. Anderson has developed a set of tools that will facilitate this work and that follow the same strategy employed previously by AK06 to produce the standard effective PSF library. 

\section{Remarks}
Most of the observations have been completed and we are starting with the analysis of the data. The results of this program will significantly improve the calibration accuracies of JWST and will provide the elements to cross calibrate HST/NICMOS, Spitzer/IRAC, and JWST/NIRcam cameras. Furthermore, the photometric observations obtained here provide the means to extend the current  standards network to fainter stars, which are likely to be more suitable for the needs of future IR instrumentation.

\acknowledgements %%% Text of acknowledgements runs on after this command.
I thank all the project collaborators at STScI: S. Arribas, E. Bergeron, R. Bohlin, S. Casertano, R. de Jong, L. Dressel, R. Guilliland, R. Hook, G. Kriss, P.  McCullough, M. McMaster, B. Mobasher, K.Noll, V. Platais, M. Regan, B. Sparks, M. Stiavelli, J. Valenti;  and at other institutions: J. Anderson (Rice University), B. Reach and J. Stauffer  (Spitzer Science Center) J. Rhoads (UA). Their contributions are  key for the success of this program.
I'm deeply grateful to J. Anderson for his help with the astrometric observations and to R. Bohlin for his continuous support and advice.  

%%% THE BIBLIOGRAPHY
%%%
%%% CONSULT SECTION 3 OF "INSTRUCTIONS FOR AUTHORS" FOR HOW TO USE NATBIB.
%%% AUTHORS ARE ENCOURAGED TO USE EITHER THE "THEBIBLIOGRAPY" ENVIRONMENT
%%% BY UNCOMMENTING (DELETING THE "%" SYMBOL) THE COMMANDS BELOW, OR BY
%%% USING THE BIBTEX ENVIRONMENT. TO FIND OUT WHICH IS APPLICABLE TO YOUR
%%% CONTRIBUTION, CONSULT THE VOLUME EDITORS FOR YOUR PROCEEDINGS.
%%%

\end{document}